\documentclass{article}
\begin{document}
\title{\bf Entropies based on fractional calculus}
\author{Marcelo R. Ubriaco\thanks{Electronic address:ubriaco@ltp.upr.clu.edu}}
\date{Laboratory of Theoretical Physics\\University of Puerto Rico\\R\'{\i}o Piedras Campus\\
San Juan\\PR 00931, USA}

\maketitle
\begin{abstract}
We propose entropy functions based on fractional calculus. We show that this new
entropy has the same properties than the Shannon entropy except  additivity.
 We show that this entropy function satisfies the
Lesche and thermodynamic stability criteria.
\end{abstract}
\vspace{0.2in}
PACS numbers: 65.40Gr\\
Keywords: entropy, fractional calculus, stability

\section{Introduction}
In the last two decades, there has been a lot of interest in generalizing
the Shannon entropy and exploring the consequences of applying these new entropies
to physics and other fields.
These entropy functions depend on an additional parameter $q$ and
become the Shannon entropy function when this parameter takes the value $q=1$.
The implications of these generalizations are not merely mathematical but
in some cases  these entropies could be  non-extensive opening the possibility for
applications to systems with long range correlations. 

The most studied 
generalization of the Shannon entropy, besides the R\'{e}nyi entropy \cite{R}, is the Tsallis entropy \cite{T}
\begin{equation}
S_q=\frac{k}{q-1}\left(1-\sum_ip_i^q\right),
\end{equation}
 where $p_i$ is the probability . The probability distribution under
the constraints $\sum_i p_i=1$ and $\sum p_i\epsilon_i=E$ is given by the function
\begin{equation}
p_i=\frac{[1-\beta(q-1)\epsilon_i]^{1/(q-1)}}{Z_q},
\end{equation}
where $Z_q$ is the partition function.
Many applications of this entropy can be found in the literature,
not only in statistical mechanics but also in other fields like
economics, biology, gravitation and high energy collisions \cite{Tsallis}. 

More recently, another  non-additive entropy was proposed \cite{S1}\cite{S2}
 according to the function
\begin{equation}
S(q)=-\sum_ip_i^q\ln p_i,
\end{equation}
leading to the probability distribution
\begin{equation}
p_i=\left(\frac{-qW(z)}{(a+bE)(q-1)}\right)^{1/(1-q)},
\end{equation}
where the variable $z$ is a function of the energy $E$ and the
parameter $q$, and $W(z)$ is the Lambert function \cite{Corless} defined by
\begin{equation}
W(z)e^{W(z)}=z  \label{lambert}
\end{equation}
 In addition,  with use of
optimization principles,
several other entropies (or measures) have been proposed \cite{Kapur}\cite{Karmeshu} to address
issues in information theory. One of these measures, is the Havrada-Charvat measure
given by the function
\begin{equation}
S(p:q)=\frac {\sum_ip_i^{\alpha}P_i^{1-\alpha}-1}{\alpha-1},
\end{equation}
which is a relative entropy that becomes the Kullback-Liebler measure for $\alpha=1$
and the Tsallis entropy for $P_i=1$. 

 In particular, in statistical mechanics, the main motivation to propose new entropies resides
in the hope that they could describe phenomena that lie  outside the scope of
the Boltzmann-Gibbs formalism. 

This paper is organized as follows.
In Section \ref{fd} we give a brief description of fractional derivatives. In Section \ref {nef}
with use of the fractional derivative
we define the new entropy, and maximize it subject to the constraint equations
$\sum_i p_i=1$ and $\sum p_i\epsilon_i=E$, and find that the probability distribution
has an exponential solution
for $q=1/2$. In Section \ref{stab} we study two criteria of stability of this entropy function,
and in Section \ref{conc} we summarize  our results.

\section{Fractional Derivatives}\label{fd}
The subject of fractional calculus  is one of the many examples of
a mathematical formalism that, in spite of its long history, it has
remained until recently practically absent in the physics literature.   L'Hopital in 1695
wondered about the meaning of $\frac{d^m f(x)}{dx^m}$ for $m=1/2$, 
and it was until the first half of the nineteen century that a precise mathematical
formulation of  fractional calculus was developed thanks to the contributions
of mathematicians like N. H. Abel, J. B. Fourier, J. Liouville and B. Riemann, among
many others \cite{Hilfer}. More recently, some applications of fractional calculus 
include the fields of anomalous diffusion, chaos, polymer science,  biophysics, 
and field theory \cite{MK}-\cite{Frederico}.
The idea behind the definition of a  fractional derivative resides in finding an  operator
that generalizes the equation 
\begin{equation}
\frac{d^nx^m}{dx^n}=\frac{m!}{(m-n)!}x^{m-n},
\end{equation}
for arbitrary $n\in R^+$, by replacing each factorial by
a gamma function
\begin{equation}
\frac{d^qx^\mu}{dx^q}=\frac{\Gamma(\mu+1)}{\Gamma(\mu-q+1)}x^{\mu-q}\;\;\;\;\;, q>0.
\end{equation}
This generalization is fulfilled by defining the operator
\begin{equation}
_aD_t^q=\left(\frac{d}{dt}\right)^n(_aD_t^{q-n}f(t)),
\end{equation}
where $n\in N$, $n>q$ and 
\begin{equation}
_aD_t^{q-n}f(t)=\frac{1}{\Gamma(n-q)}\int_a^t\frac{f(t')}{(t-t')^{(1+q-n)}}dt'.\label{lhd}
\end{equation}
Equation (\ref{lhd}) defines the fractional integral operator.  There is also
a right  Riemann-Liouville derivative, which plays an important role in
integration by parts and it is defined according to
\begin{equation}
_tD_b^q=\left(-\frac{d}{dt}\right)^n(_tD_b^{q-n}f(t)),
\end{equation}
with
\begin{equation}
_tD_b^{q-n}f(t)=\frac{1}{\Gamma(n-q)}\int_t^b\frac{f(t')}{(t-t')^{(1+q-n)}}dt'.
\end{equation}
\section{Entropy Functions}\label{nef}
The observation  that the Shannon entropy can be defined
from the equation
\begin{equation}
S=\lim_{t\rightarrow -1}\frac{d}{dt}\sum_i p_i^{-t},
\end{equation}
opened the possibility to define new entropy functions. In particular,
it was pointed out \cite{Abe} that the Tsallis entropy can be expressed in an
equivalent way as \footnote{Our Eq. (\ref{defS}) is slightly different
than the one proposed by S. Abe.}
\begin{equation}
S=\lim_{t\rightarrow -1} D_q^t \sum_i p_i^{-t}, \label{defS}
\end{equation} 
where the operator $D_q^t$ is called the Jackson \cite{Jackson} $q$-derivative
defined  as
\begin{equation}
D_q^t=t^{-1}\frac{1-q^{td/dt}}{1-q}.
\end{equation}
Since the Jackson derivative plays an important role  in the formulation of
non-commutative calculus and thus in quantum groups,  it is expected \cite{U} that
quantum groups may also play an important role in
Tsallis formalism. Other entropy functions have been defined
with the use of variations of
the Jackson $q$-derivative \cite{Borges}\cite{Johal}, a new thermostatistics based
on $q$-analysis \cite{LSS} and generalizations
of the Tsallis entropy \cite{Frank}.

Here we use the same approach as in Equation (\ref{defS}) but with the operator defined in Eq. (\ref{lhd})
with $a=-\infty$. Therefore our entropy  is given by the equation
\begin{equation}
S_q[p]=\lim_{t\rightarrow -1}\frac{d}{dt}\left(_{-\infty}D_t^{q-1}\sum_ie^{-t\ln pi}\right),
\end{equation}
where $0\leq q\leq 1$. Therefore, we need to solve the following integral
\begin{equation}
S_q[p]=\lim_{t\rightarrow -1}\frac{d}{dt}\frac{1}{\Gamma(1-q)}\sum_i\int_{-\infty}^t\frac{e^{-t'\ln p_i}}
{(t-t')^q}dt'.
\end{equation}
By defining a new variable $w=t-t'$ and with use of the definition of the $\Gamma(z)$
function
\begin{equation}
\Gamma(z)=x^z\int_0^{\infty}t^{z-1}e^{-tx} dt\;\;\;x>0\;,\;z>0
\end{equation}
we find, after taking the ordinary derivative and setting $t=-1$, that the entropy becomes
the function
\begin{equation}
S_q[p]= \sum_i (-\ln p_i)^q p_i. 
\end{equation}
It is clear  that $S[q]=\sum_i s_i(p)$ is positive, and from the condition $\frac{\partial  s_i(p)}{\partial p_i}=0$
we see that $s_i(p)$ has a maximum at $p_i=e^{-q}$ with a second derivative at this point given
by $\frac{\partial^2 s_i(p)}{\partial p_i^2}\mid_{p_i=e^{-q}}=-q^{q-1}e^q$. This maximum will be closed to $p\approx 1$ for $q\approx 0$ and close to $p\approx 0$ for $q\gg 1$. 
In particular, the binary entropy 
\begin{equation}
S_q^{bin}=p(-\ln p)^q+(1-p)(\ln (1-p))^q,
\end{equation}
has a maximum at $p=1/2$. 
In addition, as expected, this entropy is non-additive. Let ${\bf p}=(p_1,...,p_m)$ and ${\bf P}=(P_1,...,P_n)$ be two independent
probability distributions for systems $A$ and $B$ respectively. The entropy for the joint probability is given by
\begin{equation}
S_q[A,B]=\sum_{i=1}^m\sum_{j=1}^n p_iP_j\left(-\ln(p_iP_j)\right)^q.\label{Smn}
\end{equation}
A simple calculation shows that Eq. ({\ref{Smn}) becomes the infinite series
\begin{equation}
S_q[A,B]=S_q[A]+\sum_{i=1}^m\sum_{j=1}^n\sum_{k=1}\left(_k^q\right)p_iP_j(-\ln p_i)^k(-\ln P_j)^{q-k},
\end{equation}
such that exchanging $P\leftrightarrow p$ we can symmetrize this equation leading to
\begin{equation}
S_q[A,B]=\frac{1}{2}\left\{S_q[A]+S_q[B]+\sum_{k=1}\left(_k^q\right)\left[S_k[A]S_{q-k}[B]+S_{q-k}[A]S_k[B]\right]\right\}\label{add}.
\end{equation}
A simple check shows that we recover the additive property by taking the $lim_{q\rightarrow 1}S_q[A,B]$ where $S_0[p]=1$ and $S_1[p]=S_{Shannon}$. 

As is well known, the probability distributions can be obtained by maximizing
the corresponding entropy function subject to some constraint equations. Therefore,
we need to maximize
\begin{equation}
{\sl L}=S_q[p] +\alpha\left(1-\sum_i p_i\right)+\beta\left(E-\sum_i p_i\epsilon_i\right),
\end{equation} 
such that setting $\frac{d{\sl L}}{dp_j}=0$, leads to the equation
\begin{equation}
(-\ln p_j)^q- q (-\ln p_j)^{q-1}=\alpha+\beta \epsilon_j.\label{dL}
\end{equation}
For the particular case of $q=1/2$ there is an exact solution to Eq. (\ref{dL}) with
\begin{equation}
p\propto e^{(-1/2)(\Omega^2+\sqrt{\Omega^2 +2} \;\Omega)},
\end{equation}
where $\Omega=\beta\epsilon+\alpha$. 

\section{Stability}\label{stab}
In this section we investigate the stability properties of the new entropy for the
cases of Lesche and thermodynamic stability criteria.

 \subsection{Lesche stability}

In a seminal article \cite{Lesche1}\cite{Lesche2} Lesche proposed a stability criterion to study
the stability of the R\'{e}nyi entropy function. He showed that the  R\'{e}nyi
entropy is unstable for every value of the $q$ parameter with the exception of $q=1$.
Therefore, under this stability criterion the Shannon entropy is stable.  The main
motivation of this type of stability is to check whether an observable changes
appreciably when the probability assignments $p$ on a set of n microstates is
perturbed by an infinitesimal amount $\delta p$. This criteria has already been applied \cite{Abe1}-\cite{O} to some generalizations of the Shannon entropy. Let $p$ and $p'$ be two probability
assignments, Lesche stability requires that $\forall \epsilon>0$ we can find a $\delta>0$
such that 
\begin{equation}
\sum_{j=1}^n|p_j-p'_j|\leq \delta\Longrightarrow \frac{|S_q[p']-S_q[p]|}{S_q^{max}}<\epsilon.\label{Lesche}
\end{equation}
Starting from Eq. (\ref{Lesche}) and an expression for a generalized entropy maximized by
a probability distribution, the authors of \cite{AKS} derived a simple condition
from which Lesche stability can be addressed. In Ref. \cite{AKS} it is shown that Lesche stability is satisfied
if 
\begin{equation}
\frac{|S_q[p']-S_q[p]|}{S_q^{max}}<C\sum_{j=1}^n|p_j-p'_j|,\label{ineq}
\end{equation}
where the constant $C$ is given by
\begin{equation}
C=\frac{f^{-1}(0^+)-f^{-1}(1^-)}{f^{-1}(0^+)-\int_0^1f^{-1}(t)dt},\label{C}
\end{equation}
and the function $f^{-1}(t)$ is the inverse of the probability distribution.
Therefore if $C\neq 0$ , we can take an arbitrary $\delta$ as $\delta=\epsilon/C$
such that the criterion in Eq. (\ref{Lesche}) is fulfilled.
From Eq. (\ref{dL}) we see that in our case
\begin{equation}
f^{-1}(t)=(-\ln t)^q-q(-\ln t)^{q-1},
\end{equation}
such that replacing in Eq. (\ref{C}) we obtain
\begin{equation}
C=\lim_{t\rightarrow 0^{+}}\frac{f^{-1}(t)}{f^{-1}(t)-\Gamma(1+q)+q\Gamma(q)}=1,
\end{equation}
and taking $\delta=\epsilon$  the inequality in Eq. (\ref{ineq}) is satisfied and thus Eq. (\ref{Lesche}).
\subsection{Thermodynamic stability}
As is well known, in the Boltzmann-Gibbs formalism the concavity of the entropy, $\frac{\partial^2 S}{\partial U^2}<0$, 
is equivalent to the condition of thermodynamic stability. It has been recently shown \cite{SW}
that for the case of a non-additive entropy the property of concavity does not imply thermodynamic
stability. The study of thermodynamic stability of some  non-additive entropies can be found in \cite{SW}\cite{O}\cite{Wada}. Given an isolated system composed of two independent and identical subsystems 
in equilibrium in which there is a small amount of
internal energy transfered from one to the other, the property $\frac{\partial^2 S_q}{\partial U^2}<0$
does not necessarily implies that
\begin{equation}
S_q(U,U)>S_q(U+\Delta U,U-\Delta U),\label{stabeq}
\end{equation}
provided that the initial state is a state that maximizes the total entropy.
Inserting Equation (\ref{add}) into Equation (\ref{stabeq}) and expanding up to
$(\Delta U)^2$ after some algebra we find
\begin{equation}
0>\frac{1}{2}S''_q+\sum_{k=1}^{\infty}\left(_k^q\right)\left(\frac{1}{2}(S_{q-k}S''_k+S''_{q-k}S_k)-S'_kS'_{q-k}\right)\label{stabeq1}.
\end{equation}
In the microcanonical picture $S_q=\ln^q n$, and we can analyze the summation in
Equation (\ref{stabeq1}) by writing the functions $S'_k$ and $S''_k$ in terms of $S_k$
as follows
\begin{eqnarray}
S'_k&=&\frac{k}{n} S_{k-1}\nonumber\\
S''_k&=&-\frac{k}{n^2}\left(S_{k-1}-(k-1)S_{k-2}\right)\nonumber\\
S_kS_{q-k}&=&S_q,
\end{eqnarray}
such that replacing these identities into Equation (\ref{stabeq1}) and performing
the summations we find
\begin{equation}
0>\frac{1}{2}S''_q-\frac{2^q-1}{2n^2}\left(q S_{q-1}+(q-q^2)S_{q-2}\right)\label{thstab},
\end{equation}
which due to the fact that $S_q$ is a positive and  concave function Equation (\ref{thstab})
satisfies the  inequality for $ 0<q<1$ and therefore thermodynamic stability. A simple check shows that
the condition $0>S''(q)$ is recovered for $q=1$.
\section{Conclusions}\label{conc}
In this paper we  defined a new entropy function in the context of fractional calculus.  This new 
entropy is concave, positive definite and non-additive. Maximizing
the entropy subject to the usual constrains leads to an exponential probability distribution
for $q=1/2$. In addition, we have shown that this entropy satisfies Lesche and thermodynamic stability.
There are several issues one could address regarding the proposed entropy. In order to determine
whether the non-additive property of this entropy implies non-extensivity, it will require
to compute within this framework the correlation function and correlation length for a simple ideal gas 
and compare them with the $q=1$ case. For example, this type of calculation was performed in \cite{U1}
to check the factorization approach \cite{BDG} of Tsallis quantum statistics, and it was found
that correlations are smaller for $q\neq 1$ putting into question the crude approximation used in Ref. \cite{BDG}.
In addition, it is an open problem in what type of applications could  this entropy function
be successfully used, particularly in those fields where the Shannon entropy have presented limitations,
as for example in the formulation of algorithms for image segmentation \cite{PP}.  We will attempt to address
some of these questions in future communications. 
\subsection{Acknowledgment}
I am grateful to the reviewers for their suggestions that helped to
improve the content of the original version of this paper.

\newpage

Fig. 1:The entropy function for $q=1/2$  and the Shannon entropy ($q=1$)
\\
\\
\\
\\
\\
Fig. 2:The probability distribution $p(\Omega)$ for $q=1/2$ and $q=1$
\newpage
\begin{figure}
\input{figure1}
\end{figure}
\begin{figure}
\input{prob}
\end{figure}


\begin{thebibliography}{99}
\bibitem{Tsallis} A list of works can be found in tsallis.cat.cbpf.br/biblio.htm
\end{thebibliography}
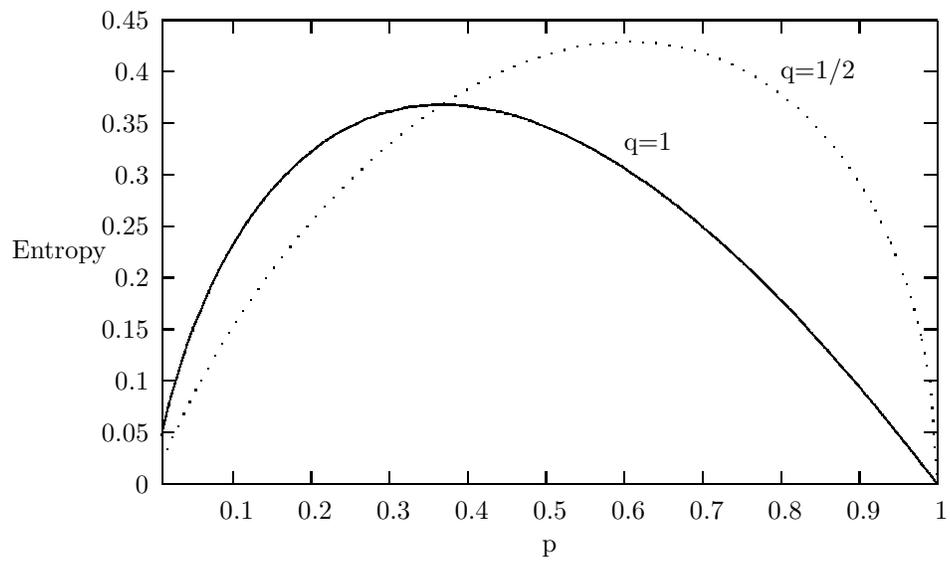
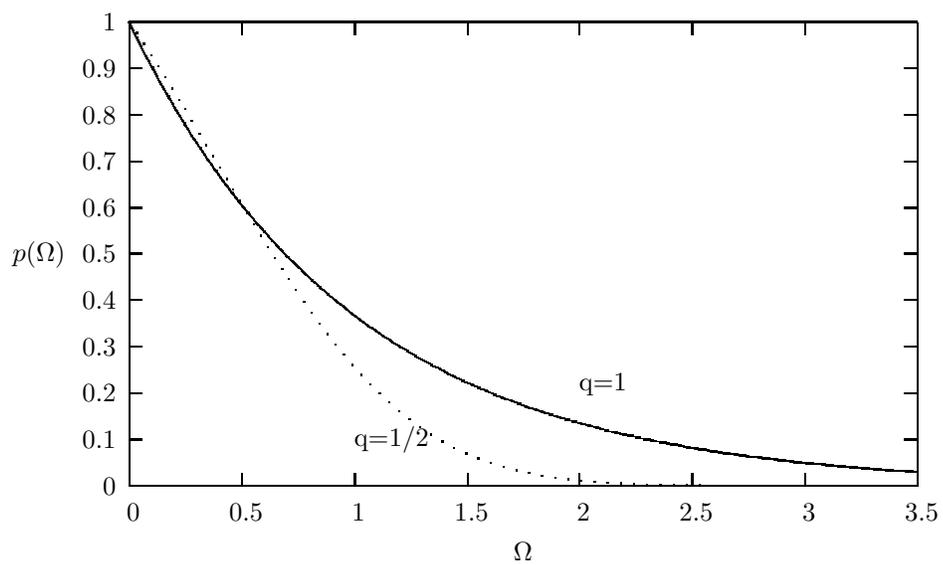
\end{document}